# A Robust eLORETA Technique for Localization of Brain Sources in the Presence of Forward Model Uncertainties

A. Noroozi, *Student Member, IEEE,* M. Ravan, *Senior Member, IEEE,* B. Razavi, R. S. Fisher, Y. Law, and Mohammad S. Hasan

*Abstract— Objective:* In this paper, we present a robust version of the well-known exact low-resolution electromagnetic tomography (eLORETA) technique, named ReLORETA, to localize brain sources in the presence of different forward model uncertainties. *Methods:* We first assume that the true lead field matrix is a transformation of the existing lead field matrix distorted by uncertainties and propose an iterative approach to estimate this transformation accurately. Major sources of forward model uncertainties, including differences in geometry, conductivity, and source space resolution between the real and simulated head models, and misaligned electrode positions are then simulated to test the proposed method. *Results:* ReLORETA and eLORETA are applied to simulated focal sources in different regions of the brain and in the presence of different levels of noise as well as real data from a patient with focal epilepsy. The results show that ReLORETA is considerably more robust and accurate than eLORETA in all cases. *Conclusion*: Having successfully dealt with forward model uncertainties, ReLORETA proved to be a promising method for real-world clinical applications. *Significance:* eLORETA is one of the most reliable techniques to study brain activity for medical applications such as determining the epileptogenic zone in patients with medically refractory epilepsy. However, eLORETA's major limitation is sensitivity to the uncertainties in the forward model. Since this problem can substantially undermine the performance of eLORETA in real-world applications where the exact lead field matrix is unknown, developing a more robust method capable of dealing with these uncertainties is of significant interest.

## I. Introduction

EXACT low-resolution electromagnetic tomography (eLORETA) is a brain source localization (BSL) technique that is widely used to localize the source of

A. Noroozi is with the Department of Digital, Technologies and Arts, Staffordshire University, Staffordshire, England, UK. M. Ravan is with the Department of Electrical and Computer Engineering, New York Institute of Technology, New York, NY, USA (correspondence e-mail: mravan@nyit.edu). B. Razavi and Robert S. Fisher are with the Department of Neurology and Neurological Sciences, Stanford University, Stanford, CA, USA. Y. Law and M.S Hasan are with the Department of Digital, Technologies and Arts, Staffordshire University, Staffordshire, England, UK.

electrical activity in the brain using electroencephalography (EEG) signals measured from a set of electrodes placed on the scalp [1], [2]. The localized brain source resulting from this so-called EEG inverse problem then can be used for a better diagnosis and treatment of mental or neurological disorders such as epilepsy [3], depression [4], and schizophrenia [5]. eLORETA is one of the most accurate methods in localizing a single focal source [6] in comparison with other distributed source estimation techniques such as minimum norm estimation (MNE) [7], weighted MNE (WMNE) [7], low-resolution electromagnetic tomography (LORETA) [8], and standardized LORETA (sLORETA) [9]. Furthermore, eLORETA shows a better performance in suppressing less significant sources and produces less blurred results in comparison with its predecessor the sLORETA technique [9], [10].

It is well-established that brain activity can be modeled by the source current distribution, which is discretized into a set of current dipoles distributed over the entire brain. Accordingly, solving the EEG inverse problem entails computing the potentials on the scalp produced by unit current dipoles in the brain using an appropriate head model, which is referred to as the forward problem [11]. The accuracy of the inverse solution depends on the employed head model [11]–[13], which in turn requires expensive magnetic resonance imaging (MRI) or computerized tomography (CT) systems, necessitating high computational complexity to incorporate the realistic geometry of the head. This can be especially problematic for studies involving a large number of participants, where taking MRIs of all participants is expensive or not possible. Typical examples include experiments carried out for gathering an EEG dataset for different applications such as emotion recognition [14], diagnosis of mental disorders [5], etc. An alternative approach is using the head models developed based on the pre-determined realistic head properties such as colin27 [15], ICBM152 [16] and New York head [17]. These head models take into account the electrical properties of the head that are included in the lead field matrix which comprises vectors of potential amplitudes received by electrodes for each active dipole source in the brain.

Previous studies have demonstrated that the performance of BSL techniques may be extremely degraded by even a slight mismatch between the true and the predetermined (simulated) lead field matrices [13], [18], [19]. On the other hand, the precise knowledge of the true lead field matrix is rarely available in practice. This is mainly because, in practical settings, many of the assumptions made about the predetermined head model, dipole sources, and the electrode array may no longer hold. In practice, when the MRI of the subject is not available, there is usually a considerable difference between the geometries of the real head and the predetermined model used for the inverse solution. This problem is referred to as geometry uncertainty in this paper. Furthermore, there is often a certain amount of error in co-registering the electrode positions as well as a considerable uncertainty in the conductivities of tissues. The difference in the location and number of dipoles between the forward and inverse models, which is referred to as discretization uncertainty in this paper, can affect the results as well [20]. In this paper, by "forward" and "inverse" models, we mean the head models used for generating EEG signals and solving the EEG inverse problem, respectively. There may also be other unknown factors that impact the calculations. As a result, a mismatch between the lead field matrix generated by the simulated forward model, and the real unknown lead field matrix is very likely. Since this mismatch can substantially degrade the performance of source localization, developing a more robust method, which is able to deal with these uncertainties adaptively, is of vital importance.

Different studies have investigated the sensitivity of the EEG source localization to forward model uncertainties including conductivity of tissues [12], [21], [22], geometry uncertainty (when MRIs are not available) [13], [23], and electrode positions [23], [24]. Analyzing a single-dipole source scenario using a five compartment (grey matter, white matter, skull, skin, and CSF) head model constructed based on T1-weighted and T2-weighted MRIs, investigators [12] reported that uncertainties in the conductivities of skin and skull have a significant effect on the accuracy of the source localization, where the simulated sources in their study were located in the somatosensory cortex. However, the inverse model's source space was not constrained to the gray matter. It was created in the intracranial space containing both gray and white matters whose conductivity uncertainties were also found to have negligible influence on the source localization but a strong effect on the strength and orientation of the reconstructed source, respectively. Similarly, the CSF conductivity uncertainty also had a trivial influence on the source reconstruction [12]. On the other hand, the distributed source estimation techniques that use cortical constraint are considerably less sensitive to skull conductivity [22]. This result was obtained using a three-layer (brain, skull, and scalp) head model constructed based on T1-weighted anatomical MRIs of 14 subjects. In one study [13], the effect of different forward model uncertainties was assessed, and it was observed that the geometry and conductivity uncertainties could cause localization errors of up to 37 mm and 22 mm, respectively. The forward model in their study was a three-layer (brain, skull, and scalp) head model derived from an MRI dataset, and the brain segmentation was used to position a large number of sources uniformly distributed across the grey matter. However, they employed a simulated three-layer isotropic spherical head model to solve the inverse problem, and the source space was created in the whole intracranial space instead of the cortex. Similarly, one study [23] reported localization errors of up to 31 mm due to the conductivity uncertainty and observed a considerable relationship between the geometry uncertainty and the error of the source localization. In their study, six head models, including 3 four-layer and 3 three-layer models, were constructed based on T1-weighted MR images of four subjects, and the simulated sources were distributed in the inner skull boundary and intracranial space for the three-layer and four-layer head models, respectively, without being constrained to the cortical area. Their simulations did not include noise and they concluded that increasing the geometry uncertainty and adding noise could lead to higher errors. They also examined the effect of electrode co-registration errors by tilting the simulated EEG montage 5 degrees to the left. In the end, they observed localization errors of up to 12 mm due to this shift in electrode positions. However, random electrode displacements have been shown not to have a significant effect on EEG source localization [24]. This result was reported using a three-layer head model derived from the MRI scan of one patient, and the simulated sources were positioned in the intracranial space with no cortical constraints.

To address the above mentioned problems, several methods have been proposed in the literature, including beamformer-based techniques [25]–[28], Bayesian approaches [29], [30], and conductivity fitting [31], [32]. In [25] a so-called diagonal loading (DL) technique is applied to the MEG source localization problem. In the DL method, the covariance matrix of the measurements is replaced with a regularized version in which a constant factor of the unity matrix is added to the initial covariance matrix. Although DL is capable of reducing the sensitivity to some extent, it leads to a trade-off between the spatial resolution of linearly constrained minimum variance (LCMV) [33] beamformer and the signal-to-noise ratio (SNR) of the output [26]. On the other hand, it is not clear how the optimal value of the DL factor can be determined given known levels of uncertainty in the lead field matrix [27]. Eigenspace-based beamformers [26], [28] are also other methods that are effective in reducing the sensitivity to modeling errors as well as measurement noise. However, these methods are susceptible to the so-called "subspace swaps" which occur at low SNR when the eigenvalues of the signal subspace cannot be distinguished from the eigenvalues of the noise subspace adequately [27]. The robust minimum variance beamformer (RMVB) [20], which inherits its structure from a technique in the signal processing field with the same name [27], [34], incorporates the uncertainty of the lead field matrix into the estimation of beamforming spatial-filter weights, which are then used to reconstruct the brain source activity. In this method, several hyper ellipsoids, *i.e.*, ellipsoids in higher dimensions, are used to define regions of uncertainty for a given nominal lead field vector. Despite the novelty of RMVB in dealing with uncertainties in the forward model, only the conductivity and discretization uncertainties are considered for the estimation of uncertainty regions in this method **Error! Reference source not found.**[20]. Moreover, the orientation of dipoles in the inverse solution is assumed to be fixed in this technique, which is also an inaccurate assumption that is very likely to be violated in practice. To overcome this problem, the authors in [35] reformulated the RMVB method by considering free orientations for dipole sources along different axes. However, the proposed method still fails to provide an efficient and accurate technique for estimating the uncertainty regions.

Bayesian approximation error (BAE) approaches have also proved to be effective in dealing with different forward model uncertainties. The BAE method is capable of alleviating the source localization error caused by unknown head geometries [29]. However, the other sources of uncertainties are not considered in that proposed method. The BAE approach can factor in the conductivity uncertainty [30]. For this purpose, the real forward model can be formulated using a standard model accompanied by an additive approximation error term to encompass the effect of the skull conductivity uncertainty. However, this approach only was tested using high SNR values of 20 dB and 30 dB, concluding that the employed BAE method becomes uncertain when the SNR drops to 20 dB. Similarly, conductivity fitting techniques also aim to improve the accuracy of the source localization by factoring in the conductivity uncertainty. In these methods, the conductivity values are fitted while estimating the location of the source simultaneously [31]. However, both Bayesian and conductivity fitting approaches still fail to address a realistic situation for which multiple uncertainties could be present simultaneously.

In this paper, we propose a novel robust version of the well-known eLORETA (ReLORETA) which does not employ any specific presumptions and can adaptively deal with different uncertainties, regardless of their nature. For this purpose, unlike the existing methods mentioned above that entail fixing some parameters such as the size of uncertainty regions or the DL factor before solving the EEG inverse problem, we consider a worst-case scenario where no individual information in setting up the model is available and let the lead field vectors be updated freely using a transformation matrix that is automatically estimated while solving the inverse problem. The proposed algorithm can be applied to any source localization problem. However, in this paper, we assess its application to simulated focal sources including one-dipole sources and focal sources with an extended activity called "extended sources", as well as a real EEG localization in an epilepsy case example. The results show that ReLORETA achieves a considerably better performance than eLORETA and remains robust in the presence of different uncertainties even when they exist simultaneously.

The rest of the paper is organized as follows. Section II presents a brief description of eLORETA method, then the proposed ReLORETA algorithm is introduced in this section as well. Section III provides the results of the proposed method for different simulated data, including single-dipole and extended sources, in the presence of lead field uncertainties as well as the results when the method is applied to the real data of a patient with epilepsy. Discussion and conclusions are given in sections IV and V, respectively.

## II. METHODS

### A. eLORETA

Assume the vector $\mathbf{x}(n) \in \mathbb{R}^{M \times 1}$ contains electric potentials measured by $M$ electrodes placed on the scalp at the $n^{th}$ time instant, then the relation between $\mathbf{x}(n)$ and the neural activity inside the brain can be described using the EEG forward equation as follows

$$\mathbf{x}(n) = \mathbf{H}\mathbf{y}(n) + \mathbf{\eta}(n) \quad (1)$$

where $\mathbf{H} \in \mathbb{R}^{M \times (3 \times K)}$ is the lead field matrix corresponding to $K$ voxels that is subject to uncertainty and can be written as

$$\mathbf{H} = [\mathbf{H}_1, \mathbf{H}_2, \mathbf{H}_3, \dots, \mathbf{H}_K] \quad (2)$$

where $\mathbf{H}_i \in \mathbb{R}^{M \times 3}$, $i \in \{1,2,\dots,K\}$, and $\mathbf{\eta}(n)$ represents the background neural activity or noise. The vector $\mathbf{y}(n) \in \mathbb{R}^{(3 \times K) \times 1}$ contains the brain source amplitudes and is an estimator of the current densities at $K$ voxels and the $n^{th}$ time instant. Given the scalp potential measurements $\mathbf{x}(n)$, the EEG inverse problem can then be translated into finding the corresponding $\mathbf{y}(n)$ that satisfies (1).

The regularized weighted minimum norm method tries to find $\mathbf{y}(n)$ by solving the following optimization problem

$$\min_{\mathbf{y}(n)} (S_{el}(n) + \alpha \mathbf{y}(n)^T \mathbf{W}\, \mathbf{y}(n)) \quad (3a)$$

$$S_{el}(n) = \|\mathbf{x}(n) - \tilde{\mathbf{x}}_{el}(n)\|^2 \quad (3b)$$

$$\tilde{\mathbf{x}}_{el}(n) = \mathbf{H}\mathbf{y}(n) \quad (3c)$$

where $\mathbf{W} \in \mathbb{R}^{(3 \times K) \times (3 \times K)}$ denotes a symmetric weight matrix, $\alpha \geq 0$ is the regularization parameter, and $\tilde{\mathbf{x}}_{el}$ represents the reconstructed EEG signals by eLORETA [1], [9]. $S_{el}(n)$ is a measure indicating how well the EEG signals are reconstructed by eLORETA at the $n^{th}$ time point. The total reconstruction error of eLORETA for $N$ time points can then be defined as

$$E_{eloreta} = \sum_{n=1}^{N} S_{el}(n) \quad (4)$$

It can be shown that the solution to this problem is linear as follows

$$\mathbf{y}(n) = \mathbf{T}\mathbf{x}(n) \quad (5)$$

where the matrix $\mathbf{T}$ is given by

$$\mathbf{T} = \mathbf{W}^{-1}\mathbf{H}^T(\mathbf{H}\mathbf{W}^{-1}\mathbf{H}^T + \alpha\mathbf{L})^+ \quad (6)$$

in which the superscript "+" denotes the Moore-Penrose pseudoinverse and $\mathbf{L}$ is the average reference operator also known as the centering matrix [9].

In the eLORETA method, the weight matrix $\mathbf{W}$ is block-diagonal where all elements are zero except for the diagonal sub-blocks denoted by $\mathbf{W}_i \in \mathbb{R}^{3 \times 3}$, $i \in (1,2,\dots,K)$. The current density at the $i^{th}$ voxel can then be computed as

$$\mathbf{y}_i(n) = \mathbf{W}_i^{-1}\mathbf{H}_i^T(\mathbf{H}\mathbf{W}^{-1}\mathbf{H}^T + \alpha \mathbf{L})^+ \mathbf{x}(n) \quad (7)$$

where $\mathbf{H}_i$ is the lead field matrix related to the $i^{th}$ voxel according to (2), and the optimum $\mathbf{W}_i$ matrices can be calculated using the following set of equations

$$\mathbf{W}_i = [\mathbf{H}_i^T(\mathbf{H}\mathbf{W}^{-1}\mathbf{H}^T + \alpha \mathbf{L})^+ \mathbf{H}_i]^{1/2}. \quad (8)$$

A simple iterative algorithm can then be used to compute the block-diagonal weight matrix $\mathbf{W}$ as follows [1]

**Step 1:** Given the averaged reference lead field $\mathbf{H}$ and a regularization parameter $\alpha \geq 0$, initialize the block-diagonal weight matrix $\mathbf{W}$ to the identity matrix.

**Step 2:** Calculate $\mathbf{W}_i$ using the symmetric square root matrix in (8).

**Step 3:** Continue until $\mathbf{W}$ converges to a final value, *i.e.,* the change in the calculated value of $\mathbf{W}$ in two consecutive iterations become negligible.

The solution for the EEG inverse problem given by (7) and the weights satisfying the system of equations in (8) define the eLORETA method.

*B. ReLORETLA*

eLORETA is a genuine solution for the EEG inverse problem with zero localization error for single-dipole sources when the given lead field matrix is accurate. In practice, the true lead field matrix is usually unknown, and obtaining an accurate result may not be simple or even possible. If the given lead field matrix is not accurate, the objective function in (3) will not converge properly, and consequently, the result of the source localization will not be reliable.

To overcome this problem, in ReLORETA technique we assume that the real lead field matrix $\tilde{\mathbf{H}}$ is a transformation of the current inaccurate lead field matrix $\mathbf{H}$ such that

$$\tilde{\mathbf{H}} = \mathbf{R}\mathbf{H} \quad (9)$$

where $\mathbf{R} \in \mathbb{R}^{M \times M}$ is the transformation matrix. The goal is to find the optimum $\mathbf{R}$, such that the resultant lead field matrix $\tilde{\mathbf{H}}$ given by (9) can make the objective function in (3) converge properly. By substituting (9) into (3), in ReLORETA the transformation matrix and the source amplitude are simultaneously calculated by solving the following optimization problem

$$\min_{\mathbf{R}}(\min_{\mathbf{y}(n)}(S_{rel}(n) + \alpha \mathbf{y}(n)^T \mathbf{W} \mathbf{y}(n))) \quad (10a)$$
$$S_{rel}(n) = \|\mathbf{x}(n) - \tilde{\mathbf{x}}_{rel}(n)\|^2 \quad (10b)$$
$$\tilde{\mathbf{x}}_{rel}(n) = \tilde{\mathbf{H}}\mathbf{y}(n) \quad (10c)$$

where $\tilde{\mathbf{x}}_{rel}(n)$ represents reconstructed EEG signals by ReLORETA. The inner optimization problem in (10a) can be solved using (3) by eLORETA. However, finding a closed form solution for the outer optimization problem is not straightforward. For this reason, we propose an iterative algorithm to simultaneously find the optimum $\mathbf{R}$ and $\mathbf{y}(n)$ as follows: in the first iteration, we assume that the existing lead field matrix is the true lead field matrix that minimizes (10a) (i. e. $\tilde{\mathbf{H}} = \mathbf{H}$), and the inner optimization problem is solved by eLORETA according to (3). In the next step, H is replaced with $\tilde{\mathbf{H}}$ from (9), and the computed $\mathbf{y}(n)$ and W are then fed into (10) to find the optimum $\mathbf{R}$ using the outer optimization problem as follows

$$\min_{\mathbf{R}}(S_{rel}(n) + \alpha \mathbf{y}(n)^T \mathbf{W} \mathbf{y}(n)) \quad (11)$$

Since $\mathbf{y}(n)$ and $\mathbf{W}$ are given by the inner optimization, the second term in (11) is constant. Given EEG measurements at $N$ time points, the optimization problem in (11) can then be reformulated as

$$\min_{\mathbf{R}} E_{reloreta} \quad (12a)$$
$$E_{reloreta} = \sum_{n=1}^{N} S_{rel}(n) \quad (12b)$$

where $E_{reloreta}$ is the reconstruction error of ReLORETA. After optimizing (12), if our initial assumption is true (*i.e.* $\tilde{\mathbf{H}} = \mathbf{H}$), then the objective function in (10a) will converge to the same value for both inner and outer optimization problems, *i.e.* (3) and (11) (please see section D in the supplementary materials for the proof). Otherwise, the lead field matrix is updated according to (9) and all steps are repeated using the updated lead field matrix. This iterative process continues until the algorithm converges. To prevent ambiguity, we use eLORETA$^j$ to show the eLORETA method used for solving (3) in the $j^{th}$ iteration. We also use similar notations for ReLORETA, $E_{reloreta}$, and $E_{eloreta}$ in the $j^{th}$ iteration and show them by ReLORETA$^j$, $E_{reloreta}{}^j$, and $E_{eloreta}{}^j$, respectively.

The convergence of ReLORETA requires that (3) and (11) converge to the same value. As the second term in both objective functions is equal, then this entails that their first terms, i. e. $S_{el}(n)$ and $S_{rel}(n)$ converge to an equal value. Given EEG measurements at $N$ time points and according to (12b) and (4), the convergence of the algorithm is then achieved when the difference between $E_{reloreta}{}^j$ and $E_{eloreta}{}^j$ converges to zero. We employ the Levenberg-Marquardt (LM) algorithm described in the next section to solve (12).

*1) LM Optimization*

LM is a well-known algorithm that is widely used to solve nonlinear least-squares problems [36]–[38]. This algorithm is able to find a solution even if it starts at a point very far from the desired minimum [36].

The LM algorithm for finding the optimum transformation matrix $\mathbf{R}$ can be described in the matrix form as follows

$$\mathbf{R}^{j+1} = \mathbf{R}^j - (\mathbf{C} + \lambda \mathbf{I})^{-1}\mathbf{D} \quad (13)$$

where $\mathbf{C}$, $\mathbf{D}$, and $\mathbf{I}$ represent the Hessian matrix, the derivative of $E_{reloreta}$ with respect to $\mathbf{R}$, and the identity matrix respectively. $j$ stands for the number of iterations, and $\lambda$ is the learning factor.

For calculating the derivative matrix $\mathbf{D}$, we first rewrite (10c) and (12b) in the matrix form as

$$\tilde{\mathbf{X}}_{rel} = \tilde{\mathbf{H}}\mathbf{Y} = \mathbf{R}\mathbf{H}\mathbf{Y} \quad (14a)$$
$$E_{reloreta} = \|\mathbf{X}_{diff}\|_F^2 \quad (14b)$$
$$\mathbf{X}_{diff} = \mathbf{X} - \tilde{\mathbf{X}}_{rel} \quad (14c)$$

where $\|.\|_F$ denotes the Forbenius norm. $\mathbf{X}$, $\widetilde{\mathbf{X}}_{rel}$, and $\mathbf{Y}$ matrices contain EEG measurements, reconstructed EEG signals by ReLORETA, and source amplitudes at $N$ time instants as follows

$$\mathbf{X} = [\mathbf{x}(1), \mathbf{x}(2), \mathbf{x}(3), \ldots, \mathbf{x}(N)] \quad (15a)$$
$$\widetilde{\mathbf{X}}_{rel} = [\tilde{\mathbf{x}}_{rel}(1), \tilde{\mathbf{x}}_{rel}(2), \tilde{\mathbf{x}}_{rel}(3), \ldots, \tilde{\mathbf{x}}_{rel}(N)] \quad (15b)$$
$$\mathbf{Y} = [\mathbf{y}(1), \mathbf{y}(2), \mathbf{y}(3), \ldots, \mathbf{y}(N)] \quad (15c)$$

Computing $\mathbf{D}$ then entails taking the derivative of a scalar function (i.e. $E_{reloreta}$) with respect to a matrix (i.e. $\mathbf{R}$) that obeys the chain rule and can be computed as follows

$$\mathbf{D} = \frac{\partial E_{reloreta}}{\partial \mathbf{R}} = \left(\frac{\partial E_{reloreta}}{\partial \mathbf{X}_{diff}}\right)\left(\frac{\partial \mathbf{X}_{diff}}{\partial \mathbf{R}}\right) \quad (16)$$

Using the matrix algebra, we have

$$\frac{\partial E_{reloreta}}{\partial \mathbf{X}_{diff}} = 2\mathbf{X}_{diff} \quad (17a)$$
$$\frac{\partial \mathbf{X}_{diff}}{\partial \mathbf{R}} = -(\mathbf{HY})^T \quad (17b)$$

If we substitute (17a) and (17b) into (16) and use (14a), then the derivative matrix $\mathbf{D}$ is given by

$$\mathbf{D} = 2(\mathbf{RHY} - \mathbf{X})(\mathbf{HY})^T \quad (18)$$

In order to calculate the Hessian matrix, we can first rewrite (10c) in the form of the output equation of a single layer neural network as follows

$$\tilde{\mathbf{x}}_{rel}(n) = f(\mathbf{Rz}(n)) \quad (19)$$

where $\mathbf{z}(n)$ is given by

$$\mathbf{z}(n) = \mathbf{Hy}(n) \quad (20)$$

and $f$ is a linear activation function such that

$$f(\mathbf{Rz}(n)) = \mathbf{Rz}(n). \quad (21)$$

The Hessian matrix can then be calculated using the method proposed in [39] as follows

$$\mathbf{C} = f''(\tilde{\mathbf{x}}_{rel}(n))\frac{\partial S_{rel}(n)}{\partial \tilde{\mathbf{x}}_{rel}(n)} + (f'(\tilde{\mathbf{x}}_{rel}(n)))^2 \frac{\partial^2 S_{rel}(n)}{\partial \tilde{\mathbf{x}}_{rel}(n)^2} \quad (22)$$

where $f'$ and $f''$ denote the first and second derivatives of $f$. In the proposed method $f$ is linear, so we have $f' = 1$ and $f'' = 0$. By substituting these values in (22), the Hessian matrix will then be computed as

$$\mathbf{C} = \mathbf{I} \quad (23)$$

*2) Proposed Algorithm*

Using (13), (18), and (23), we can update the transformation matrix $\mathbf{R}$ in each iteration, and then, it can be fed into the eLORETA$^j$ to find the source amplitudes, i.e., the $\mathbf{Y}$ matrix. The algorithm will continue iteratively until it converges. If we define the differential reconstruction error (DRE) for iteration $j$ as

$$DRE^j = \|E_{reloreta}{}^j - E_{eloreta}{}^j\| \quad (24)$$

then the convergence of the algorithm entails that $DRE^j$ or the difference between two values of *DRE* in two consecutive iterations drops below a certain threshold value $\varepsilon$. However, to make the threshold value independent of the unit of measurements, the *DRE* value in each iteration can be normalized as follows

$$NDRE^j = \frac{\|DRE^j - DRE^{j-1}\|}{\max(DRE) - \min(DRE)}, \quad j \geq 2 \quad (25)$$

where

$$DRE = [DRE^1, DRE^2, \ldots, DRE^j] \quad (26)$$

and $NDRE^j$ denotes normalized $DRE^j$. Finally, the convergence of the algorithm is achieved when $NDRE^j \leq \varepsilon$. The proposed ReLORETA algorithm is summarized in the flowchart shown in Fig. 1 and can be described as follows:

**Step 1**: Initialize the matrices $\mathbf{R}$, $\mathbf{H}$, and set the threshold $\varepsilon$.
**Step 2**: Given the measurement matrix $\mathbf{X}$ and the lead field matrix, find the source amplitude matrix $\mathbf{Y}$ using eLORETA$^j$.
**Step 3**: Update the transformation matrix $\mathbf{R}$ using (13), (18), and (23).
**Step 4**: Calculate *NDRE* using (24) and (25). If $NDRE^j \leq \varepsilon$ then terminate the algorithm, else go to the next step.

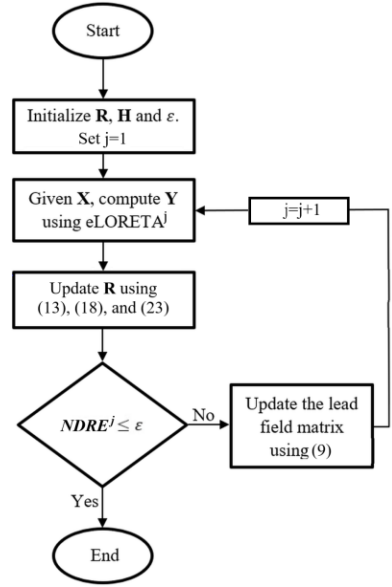

Fig. 1. Flowchart of the proposed ReLORETA algorithm.

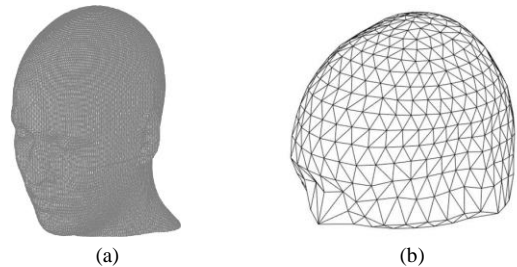

Fig. 2. Different head models used for simulations: (a) the New York head model for generating EEG signals and (b) the head model based on the Colin27 MRI to solve the inverse problem.

**Step 5**: Update the lead field matrix using (9), replace **H** with **H̃**, and go to step 2.

After the algorithm reaches the desired threshold and stops, **Y** can then be used as the estimated source amplitudes.

## III. RESULTS

### A. Simulation Settings

In order to assess the performance of the proposed algorithm in a realistic situation, we considered two scenarios. In the first scenario, called the worst-case scenario, we employed ReLORETA to localize simulated focal sources in different regions of the brain while taking into account different sources of uncertainties, including the geometry of the head model, misalignment of electrodes, discretization and conductivity uncertainties simultaneously. For this purpose, we used two different head models of different shapes for generating the EEG signals and solving the EEG inverse problem as shown in Fig. 2(a) and Fig. 2(b), respectively. As shown in this figure, the New York head was employed to generate the simulated EEG data, while a head model based on the well-known Colin27 MRI was used to solve the inverse problem. For constructing the forward model, the segmentation data provided in [17] were used to create a hexahedral mesh for six tissues, including gray and white matter, cerebrospinal fluid (CSF), skull, skin, and air using the MATLAB FieldTrip toolbox [40]. The FieldTrip-SimBio pipeline [41] was then used to construct the forward head model using the different tissues and their respective conductivities. After creating the head model, the electrode positions were aligned with the surface of the head, and then dipoles were uniformly spread within the gray matter with an orientation orthogonal to the surface of the cortex. The smallest distance between the source space dipoles and the inner skull boundary also was 3 mm. In the end, the finite element method (FEM) was employed to compute the lead field matrix using the FieldTrip toolbox. For the inverse head model, the Colin27 MRI was first segmented into three different tissues: scalp, skull, and brain, and then three meshes were created at the borders of these tissues represented by points (vertices) connected in a triangular way. The number of vertices for the scalp, skull and brain was 1000, 2000, and 3000, respectively. These meshes, together with their respective conductivities, were then fed into FieldTrip to construct the inverse head model. In the next step, the electrode positions were aligned with the surface of the head and dipoles were created in the intracranial space without being constrained to the cortex. The symmetric boundary element method (BEM) presented in the OpenMEEG software [42] was finally used to compute the lead field matrix. The forward and inverse head models were co-registered using a 3D transformation including rotation, translation, and scaling according to the method described in [43]. In this method, the transformation is calculated between 3 or more noncoplanar landmark points in the inverse model and their corresponding points in the forward model.

The simulated EEG data were generated for 20 electrodes placed on the scalp according to the 10-20 placement system [44] using the Simulating Event-Related EEG Activity (SEREEGA) package [45]. In this package, the event-related brain potentials (ERPs) are first generated for each active dipole by centering a normal probability density function around the indicated latency with the given width. This is then scaled to the indicated amplitude. To simulate noise, in the first step, Brown noise [46] is added to the simulated ERP signals, and the noisy ERP signals are projected to the scalp using the forward lead field matrix and orientations of dipoles. At this stage, a final layer of noise is added to the generated scalp data to simulate sensor noise. This is uniform, temporally and spatially uncorrelated white noise. To improve the effective peak to background EEG signal ratio, multiple trials from repeated stimuli were averaged to extract the final EEG waveforms. Specifically, for each EEG data, we employed 10 trials with a sampling frequency of 500 Hz, where each trial had a length of 200 samples, corresponding to a time sequence which may be denoted using MATLAB notation as [0: 2: 399] ms.

To simulate the error in the misalignment of electrodes, we slightly changed the locations of electrodes in comparison with the standard 10-20 system by tilting the EEG set up to the left by 5 degrees in the forward model. As for source discretization, we used a 2 mm grid for the forward model, while a coarser grid of 10 mm was used for the inverse model. Furthermore, for the forward model, we set the conductivities of the gray and white matters, CSF, skull, skin, and air to 0.8 S/m, 0.4 S/m, 2.2 S/m, 0.0290 S/m, 0.25 S/m, and 2.5e-14

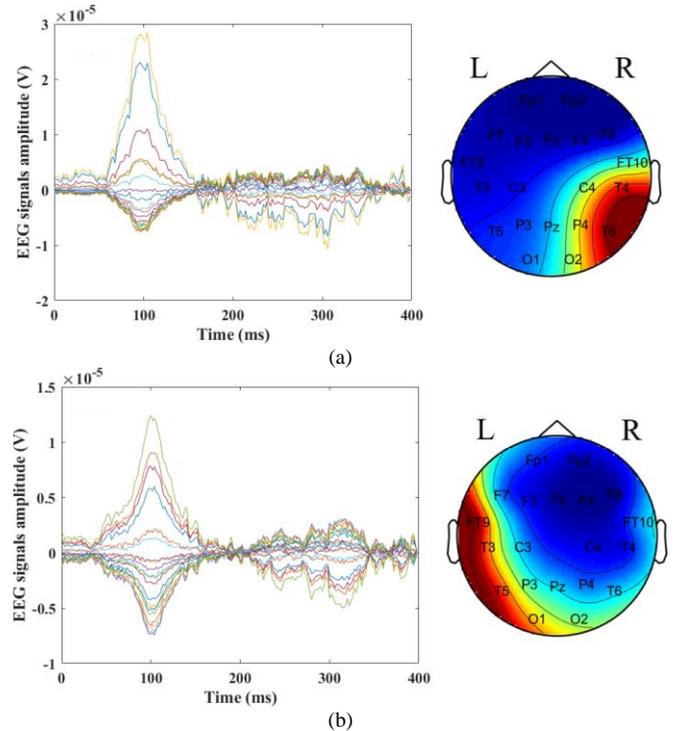

Fig. 3. Simulated EEG signals and their respective topographic distributions for two different single-dipole sources located at (a) the MNI coordinates of [50, -20, -12] mm and (b) the MNI coordinates of [-24, -14, -20] mm, corresponding to right Brodmann area 21 (middle temporal gyrus) and the left hippocampus, respectively.

S/m, respectively. However, the conductivities of the scalp, the skull, and the brain in the inverse model were set to 0.33 S/m, 0.0041 S/m, and 0.33 S/m respectively. All of the conductivity values were chosen according to the conductivity values and ratios of the real human head tissues reported in [47]. We deliberately set the conductivity of the skull in the inverse model to a value different from the conductivity of the skull in the forward model to incorporate the effect of this uncertainty into our simulations.

In all cases, the EEG data were generated in the presence of high and low levels of noise associated with SNR values of 5 dB and 20 dB, respectively. The SNR was calculated in dB according to [35] as follows:

$$SNR = 10\log_{10}(\frac{\text{tr}(CR_s)}{\text{tr}(CR_n)}) \quad (27)$$

where $CR_s \in \mathbb{R}^{M \times M}$ is the scalp ERP signal covariance matrix and $CR_n \in \mathbb{R}^{M \times M}$ is the background EEG noise covariance matrix. In this paper, the $CR_n$ matrix is calculated using EEG data when the related source was not active.

In all simulations, the regularization parameter $\alpha$ was set to 0.05 which is FieldTrip's default value (see the supplementary data for further details of tuning $\alpha$). To set the threshold value $\varepsilon$, a three-step process was implemented as follows: in the first step, 20 EEG data from 20 different sources, uniformly distributed within the gray matter, were generated. In the second step, we set $\alpha = 0.05$ and ran ReLORETA 60 iterations for all 20 sources and calculated the $NDRE$ values for each iteration according to (25). In the last step, we observed that in all cases when $NDRE^j$ falls below a certain value named $NDRE_{cutoff}$, the localization error of ReLORETA converges to its final value. So, we finally set $\varepsilon = NDRE_{cutoff}$. Using this process, the values of $\varepsilon$ for both single-dipole and extended sources were set to 0.005.

The transformation matrix **R** was initialized to the identity matrix, and the initial lead field matrix was computed using the inverse head model according to the 10-20 electrode placement system. Finally, localization error for single-dipole sources was simply calculated as the Euclidean distance between the estimated and the true source positions. For extended sources, the distance between the center of the predicted source, which is the dipole with the greatest moment, and the center of the true source structure was considered as the source localization error. In all simulations in this paper, unconstrained source orientations were used for solving the inverse problem.

*B. Single-Dipole Source Results*

In order to evaluate the performance of ReLORETA in dealing with the underlying source activity being modeled as a single dominant dipole, we simulated 60 single-dipole sources in different lobes and depths of the gray matter where the depth of each source was measured according to its distance from the nearest electrode on the scalp. The distances between the most superficial source and the nearest electrode on the scalp and the most superficial source and the inner surface of the skull were approximately 14 mm and 5 mm, respectively. Fig. 3(a) and Fig. 3(b) show two generated EEG signals with a SNR of 5 dB and their respective interpolated topographic distributions for a superficial source (*i.e.*, near the surface of the cortex) and a deep source located at MNI coordinates of [50, -20, -12] mm and [-24, -14, -20] mm. These coordinates reside in the temporal lobe and correspond to the right middle temporal gyrus (Brodmann area 21; superficial source) and the left hippocampus (deep source), respectively. The temporal lobe is specifically of important relevance to neurological disorders and is commonly involved in medically refractory epilepsy [48]. Moreover, many epileptic seizures begin in the hippocampus, as 50-75% of patients with epilepsy who have autopsies had damage to the hippocampus [49]. The localization results of eLOREATA and ReLORETA related to these EEG signals are also illustrated in Figs. 4 and 5, respectively. The estimated source locations and their corresponding localization error are shown in Table I, with ReLORETA showing substantially better performance than eLORETA for both source locations.

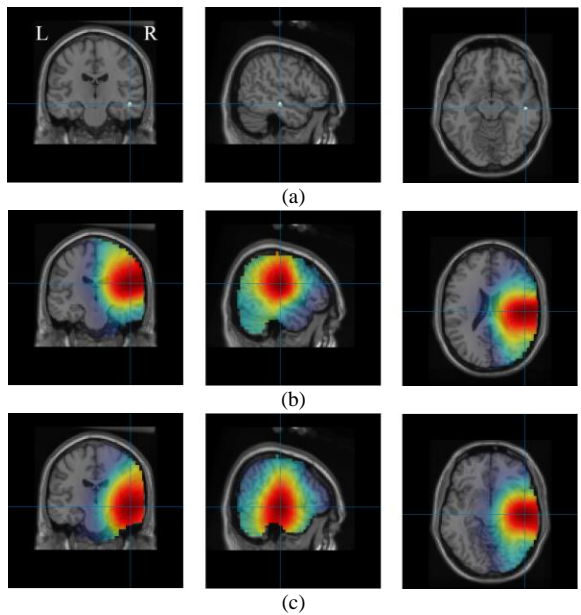

Fig. 4. (a) true source location (simulated) and localization results of (b) eLORETA and (c) ReLORETA, for the EEG data shown in Fig. 3(a).

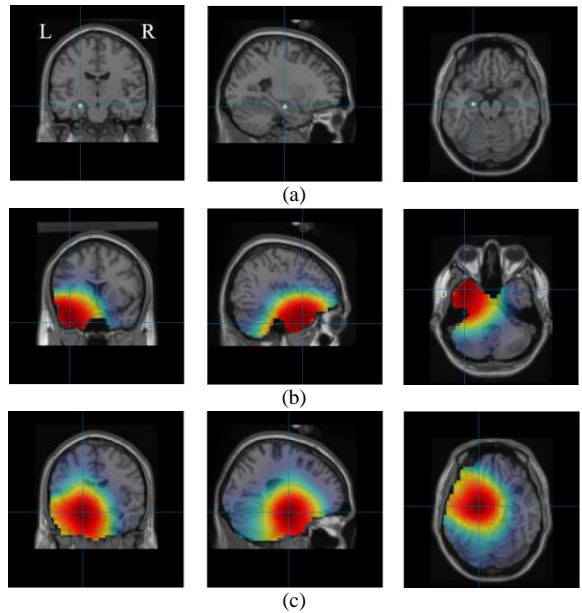

Fig. 5. (a) true source location (simulated) and localization results of (b) eLORETA and (c) ReLORETA, for the EEG data shown in Fig. 3(b).

TABLE I
SOURCE LOCALIZATION ERROR FOR SIMULATED SINGLE-DIPOLE SOURCES USING ELORETA AND THE PROPOSED RELORETA TECHNIQUES

| Source area | Source location (mm) | Estimated location using eLORETA (mm) | Estimation error using eLORETA (mm) | Estimated location using ReLORETA (mm) | Estimation error using ReLORETA (mm) |
|---|---|---|---|---|---|
| Right Brodmann area 21 | [50 -20 -12] | [50 -20 20] | 32 | [50 -20 -10] | 2 |
| Left Hippocampus | [-24 -14 -20] | [-40 10 -40] | 35.09 | [-20 -10 -20] | 5.65 |

To further evaluate the accuracy of the proposed ReLORETA method, the overall errors for all 60 simulated sources in the presence of two levels of noise were calculated for both methods and are shown in Fig. 6. The results clearly demonstrate that the proposed ReLORETA algorithm improves the overall accuracy of the source localization considerably. In terms of noise levels, both methods performed similarly and their accuracy slightly decreased when a high level of noise was applied. However, the error of ReLORETA still remains below 13 mm in the worst case. It is worth mentioning that we deliberately put the simulated sources in the voxels of the forward model which their equivalent voxels did not exist in the inverse model, such that the distance between the nearest dipole to the source center and the true location of the source in the inverse model was at least 2 mm. This is the reason for the minimum error of 2 mm in the outputs due to discretization.

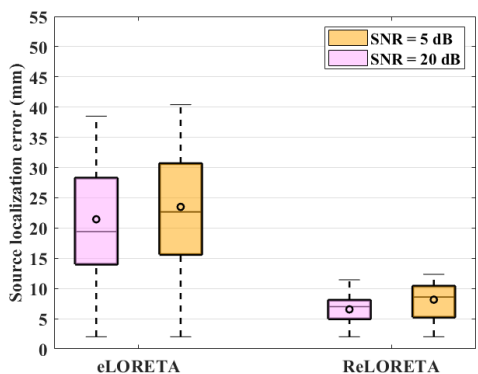

Fig. 6. The overall errors of source localization for simulated single-dipole sources. For each box, the central horizontal line indicates the median, and the top and bottom edges indicate the $75^{th}$ and $25^{th}$ percentiles, respectively. The upper and lower whiskers extend to the most extreme data points not considered outliers. The symbol "o" depicts the average value of each box.

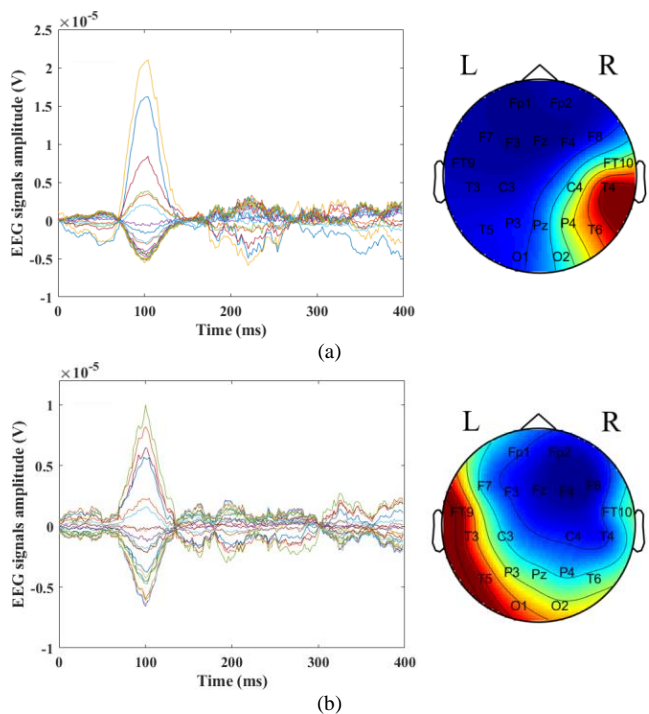

Fig. 7. Simulated EEG signals and their respective topographic distributions for two extended sources located at (a) the MNI coordinates of [50, -20, -12] mm and (b) the MNI coordinates of [-24, -14, -20] mm, corresponding to right Brodmann area 21 (middle temporal gyrus) and the left hippocampus, respectively.

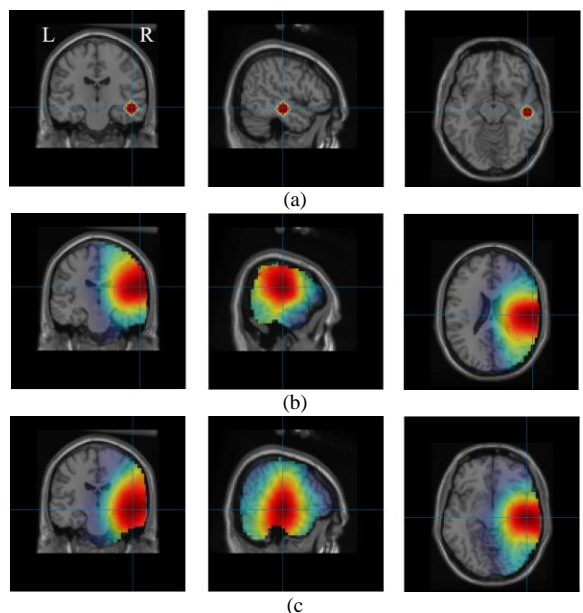

Fig. 8. (a) true source location (simulated) and localization results of (b) eLORETA and (c) ReLORETA for the EEG data shown in Fig. 7(a).

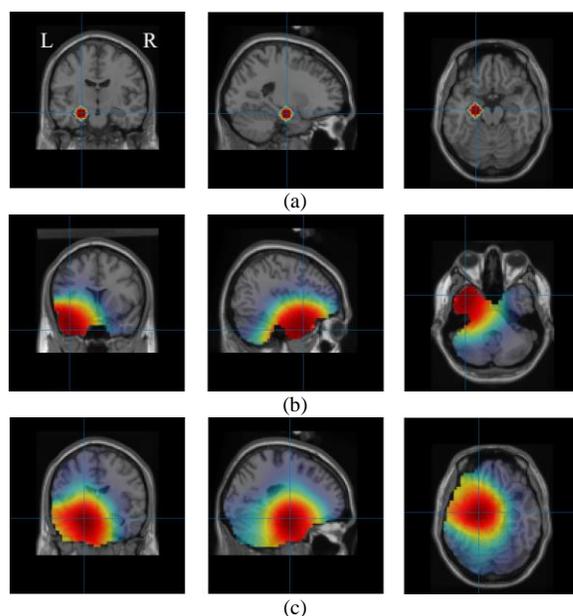

Fig. 9. (a) true source location (simulated) and localization results of (b) eLORETA and (c) ReLORETA, for the EEG data shown in Fig. 7(b).

*C. Extended Source Results*

For evaluating extended sources, we simulated 60 different sources at the same locations of the single-dipole sources while each source contained 33 active dipoles such that the total moment of dipoles was set to be the same as single-dipole sources. For this purpose, we placed a dipole in the center of the source and then spread the rest of the dipoles within a 10 mm radius. Figs. 7(a) and (b) show the simulated EEG signals for two extended sources with centers located at the same locations as two single-dipole sources: right middle temporal gyrus (Brodmann area 21) and left hippocampus, respectively. The source localization results can be found in Figs. 8 and 9 and Table II. Similar to performance for the single-dipole sources, the ReLORETA outputs are still considerably more accurate than the eLORETA outputs for both areas. The overall errors of source localization for all 60 extended sources are also shown in Fig. 10. While the performance of both methods slightly decreased in comparison with their estimations for single-dipole sources, the accuracy of ReLORETA still remained stable with average errors of 6 mm and 8 mm for low and high levels of noise, respectively. It should be also noted that the nearest dipole to the true location of the source center in the inverse model is displaced by 2 mm because of discretization. By deducting this 2 mm error from the overall error, we can conclude that on average only less than 5 mm of the total error is due to the ReLORETA output. This is while, as mentioned before, the radii of the simulated sources in the forward model are also 10 mm. This means that, even when ReLORETA estimates a source with the mentioned 5 mm bias, the estimated source value still remains inside the boundary of the source with respect to its central dipole. However, as mentioned before, the central dipole of the source in the inverse model is always at a 2 mm distance from the true location of the source center.

In the worst-case scenario discussed above, we assumed no individual information, including no available MRI of the subject. However, for a number of source localization problems, including epilepsy source localization, MRIs of patients are available. For this reason, in the second scenario we assumed that the MRI of the subject is available and the geometry of the inverse and forward models are the same. For this purpose, we used a six-shell New York head as the forward model, and a simplified three-shell and a six-shell New York head model were employed for the source localization. Four simulations were then carried out in this scenario as follows: in the first simulation we distorted the forward model by the three remaining uncertainties including geometry and conductivity uncertainties as well as errors in electrode positions. In the rest of the simulations, the forward model was distorted only by one source of uncertainty individually in each simulation while other sources of uncertainties were removed from the forward model.

After carrying out the simulations, the following results were obtained: in all cases, we observed that ReLORETA results were considerably more accurate than eLORETA. Also, the accuracy of both methods only decreased negligibly when using the three-shell model, while eLORTEA's accuracy decreased more than ReLORETA. In the first simulation, when simulating the three remaining uncertainties simultaneously, we observed that the overall error of both methods decreased compared with the worst-case scenario. In the rest of the simulations, misalignment of electrodes, conductivity uncertainties, and discretization uncertainties had the greatest impacts on eLORETA accuracy for single-dipole source localizations by causing average errors of 12 mm, 11 mm, and 8 mm, respectively. The maximum errors of eLORETA for these uncertainties were also 32 mm, 24 mm, and 18 mm respectively. The same uncertainties applied to extended sources caused average errors of 15 mm, 13 mm, and 9 mm, and maximum errors of 40 mm, 30 mm, and 26 mm, respectively for eLORETA. On the other hand, for ReLORETA, the discretization uncertainty, misalignment of electrodes, and conductivity uncertainty had the greatest impacts on the results with average errors of 7 mm, 6 mm, and 2 mm respectively for single-dipole sources. The maximum errors of ReLORETA for these uncertainties were also 9 mm, 14 mm, and 10 mm, respectively. For extended sources, the same uncertainties caused average errors of 7 mm, 6 mm and 5 mm, and maximum errors of 11 mm, 14 mm, and 14 mm, respectively. To evaluate sensitivity to electrode placement errors, we displaced electrode positions randomly and observed that the resulting error was considerably lower than when the EEG set up was tilted. All of the mentioned results are related to simulations with a 5 dB noise. The complete results of this scenario, the noise sensitivity of eLORETA and ReLORETA, and the effect of the number of samples on the results are reported in the supplementary materials.

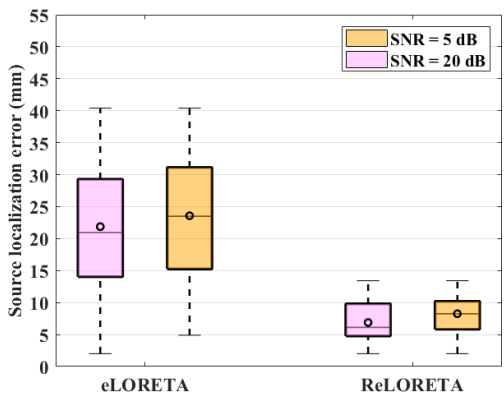

Fig. 10. The overall errors of source localization for simulated extended sources. On each box, the central horizontal line indicates the median, and the top and bottom edges indicate the $75^{th}$ and $25^{th}$ percentiles, respectively. The upper and lower whiskers extend to the most extreme data points not considered outliers. The symbol "o" depicts the average value of each box.

TABLE II
SOURCE LOCALIZATION ERROR FOR SIMULATED EXTENDED SOURCES USING ELORETA AND THE PROPOSED RELORETA TECHNIQUES

| Source area | Source center location (mm) | Estimated location using eLORETA (mm) | Estimation error using eLORETA (mm) | Estimated location using ReLORETA (mm) | Estimation error using ReLORETA (mm) |
|---|---|---|---|---|---|
| Right Brodmann area 21 | [50 -20 -12] | [60 -20 20] | 33.52 | [50 -20 -10] | 2 |
| Left Hippocampus | [-24 -14 -20] | [-40 10 -40] | 35.09 | [-30 -20 0] | 5.65 |

### D. Real Data Results

In this section, we present the result of the proposed ReLORETA algorithm for localization of the underlying brain source (*i.e.*, epileptiform discharge) in a patient with epilepsy where the related data contained EEG signals of 20 channels corresponding to the same 20 electrodes used for the simulations in sections II and III. On scalp EEG, the discharge was located on the left, centered at electrode F7, which classically represents activity predominantly from the left anterior temporal region. We employed the same inverse model we used in the simulations where the conductivities of the scalp, the skull, and the brain were set to 0.33 S/m, 0.0041 S/m, and 0.33 S/m, respectively. The regularization parameter $\alpha$ and the threshold $\varepsilon$ were set to 0.05 and 0.005, respectively similar to the simulations. The true location of the source was near the amygdala and hippocampus boundary, with a center located at the MNI coordinates of [-20 -5 -20] mm approximately. These coordinates are based on the location of seizures captured using intracranial recordings with depth electrodes. More specifically, the patient had depth electrodes implanted in the left and right amygdala, and hippocampus. However, the true location of the source can never be known with a zero error as an infinite or even a large number of electrodes cannot be implanted.

Fig. 11 shows the EEG measurements as well as their interpolated topographic distribution where each trace is averaged over 13 measured spikes. The localization results of eLORETA and ReLORETA can also be found in Fig. 12 and Table III. The first noteworthy point is that errors of both methods nearly fall into the range that had previously been predicted in our simulations for extended sources, *i.e.*, Fig.10, with localization errors of 41.53 mm and 20.61 mm for eLORETA and ReLORETA, respectively. This clearly shows that the most important sources of uncertainties were successfully included in the forward model used for simulations such that the accuracies of both methods correspond to the simulation results. Moreover, the accuracy of ReLORETA is considerably better than eLORETA, underscoring the robustness of the proposed algorithm with real-life data as well. The EEG reconstruction error of both methods as well as the convergence of the proposed algorithm for the real EEG data are shown in Fig. 11 have been provided in the supplementary materials.

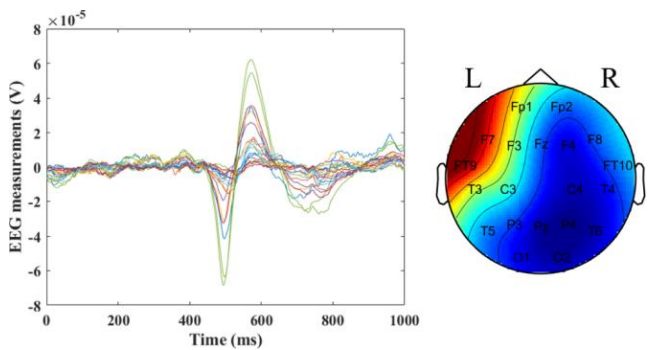

Fig. 11. Superimposed real EEG signals from an actual patient based on the 20 electrodes used for source localization and the related topographic distribution. Each trace has been averaged over 13 measurements (epileptiform discharges).

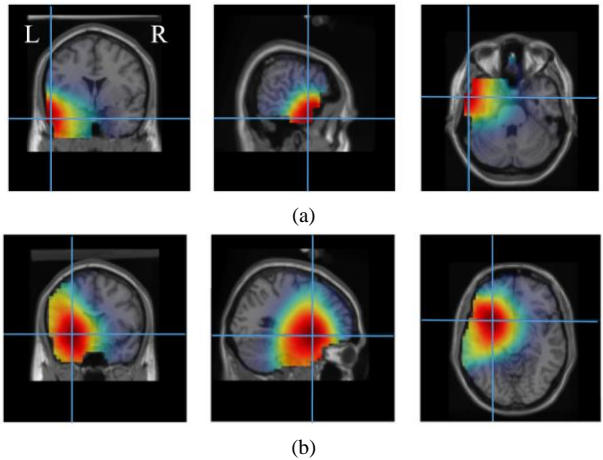

Fig. 12. Source localization results for the real EEG measurements depicted in Fig. 12. The actual source location is around the left amygdala-hippocampal junction centered at MNI coordinates of [-20 -5 -20] approximately. (a) eLORETA output and (b) ReLORETA output.

## IV. DISCUSSION

In this paper we propose a robust version of eLORETA method, named ReLORETA, for the localization of brain sources while the forward model is distorted by uncertainties. To the best of our knowledge, this is the first study proposing a method capable of dealing with major sources of uncertainties simultaneously and without any individual information in setting up the model. To examine the performance of the proposed method, we first simulated a worst-case scenario where different sources of forward model uncertainties including the conductivity uncertainty, error in co-registration of the electrodes, as well as differences in the geometry and source space resolution between the forward and inverse models, were present simultaneously. The simulation results in this scenario show that ReLORETA provides substantially more accurate results than eLORETA in all cases of single-dipole and extended sources. In the second scenario we simulated different uncertainties both simultaneously and individually while the geometries of the forward and inverse models were the same, which is equivalent to a situation where the MRI of the subject is available. For this purpose, a six-shell New York head model was used to generate EEG signals, and a simplified three-shell and a six-shell New York head model were used for the source localization. The results of ReLORETA in this scenario were also substantially more accurate than eLORETA. Furthermore, ReLORETA demonstrated a superior performance in dealing with different individual uncertainties in comparison with previous studies [20], [30], [31], [35]. The accuracy of both methods only negligibly decreased when using the three-shell model compared with the six-shell model, while eLORETA's performance was affected more than ReLORETA's performance. These results corroborate the previous findings [12] that gray matter, white matter, and CSF conductivities have a negligible effect on the source localization since their omission did not affect the localization results considerably. However, the accuracy of both methods in the second scenario increased compared with the worst-case scenario confirming the results of other studies [13], [23] that reported increased localization error due to the geometry uncertainty. This also suggests that using a head model based on the MRI of the subject can substantially improve the localization results and should be a priority. According to the results, error in co-registration of electrodes can seriously degrade the source localization results, and for both eLORETA and ReLORETA, it was one of the major sources of uncertainties, which confirms the previous findings [23]. The proposed method was also applied to real EEG data from a patient with epilepsy, and the localization accuracy for both eLORETA and ReLORETA in this case mirrored the results of the simulated data. ReLORETA remained robust in dealing with the real data and showed better performance than eLORETA. The ability of ReLORETA to show reasonable localization accuracy with limited numbers of electrodes may be of great value in clinical settings where applying high numbers of electrodes is not practical.

TABLE III
SOURCE LOCALIZATION ERROR FOR REAL DATA USING ELORETA AND RELORETA TECHNIQUES

| Approximate source center location (mm) | Estimated location using eLORETA (mm) | Estimation error using eLORETA (mm) | Estimated location using ReLORETA (mm) | Estimation error using ReLORETA (mm) |
|---|---|---|---|---|
| [-20 -5 -20] Left Amygdala-Hippocampal Junction | [-60 0 -30] Left Temporal Pole (Brodmann area 38) | 41.53 | [-30 10 -10] Anterior Insula | 20.61 |

All simulations and experiments in this paper were performed on a computer with 8 GB of RAM and an Intel Core i7-8565U 1.80 GHz processor using MATLAB R2020a. We observed that, in all cases, ReLORETA successfully converged in less than 60 iterations and usually closer to 25. Each iteration required an average time of 2 seconds to complete resulting in 50 seconds for the algorithm to converge. This time is reasonably short for an iterative algorithm, especially for the applications in which the source localization is performed offline.

## V. Conclusion

In this paper, we presented a robust version of eLORETA method called ReLORETA to deal with different uncertainties in the forward modeling that affects the source localization accuracy. For this purpose, we first assumed that the true lead field matrix is a transformation of the existing inaccurate lead field matrix, and then we proposed an iterative algorithm, in which we took advantage of the Levenberg-Marquardt optimization, for estimating this transformation accurately. In order to assess the accuracy of the proposed method, we first created a realistic forward model by taking into account different sources of uncertainties, including the conductivity and geometry of the head model, misalignment of electrodes, and discretization of source space. The ReLORETA algorithm was then applied to single-dipole and extended sources generated by the realistic head model as well as real data from a patient with epilepsy. The results showed that the proposed ReLORETA algorithm gives considerably better results for both simulated and real data in comparison with eLORETA and can be a reliable method for practical applications.

## References

[1] R. D. Pascual-Marqui, "Discrete, 3D distributed, linear imaging methods of electric neuronal activity. Part 1: exact, zero error localization," *ArXiv07103341 Math-Ph Physicsphysics Q-Bio*, Oct. 2007, Accessed: Jul. 10, 2021. [Online]. Available: http://arxiv.org/abs/0710.3341

[2] M. Iwai, N. Seihou, K. Kobayashi, and W. Sun, "Evaluation of Sensor and Analysis area in the Signal Source Estimation by Spatial Filter for Magnetocardiography," *IEEE Trans. Magn.*, pp. 1–1, 2021, doi: 10.1109/TMAG.2021.3083329.

[3] H. Rajaei *et al.*, "Dynamics and Distant Effects of Frontal/Temporal Epileptogenic Focus Using Functional Connectivity Maps," *IEEE Trans. Biomed. Eng.*, vol. 67, no. 2, pp. 632–643, Feb. 2020, doi: 10.1109/TBME.2019.2919263.

[4] Y. Zhang, T. Gong, S. Sun, J. Li, J. Zhu, and X. Li, "A functional network study of patients with mild depression based on source location," in *2020 IEEE International Conference on Bioinformatics and Biomedicine (BIBM)*, Dec. 2020, pp. 1827–1834. doi: 10.1109/BIBM49941.2020.9313315.

[5] K. Masychev, C. Ciprian, M. Ravan, J. P. Reilly, and D. MacCrimmon, "Advanced Signal Processing Methods for Characterization of Schizophrenia," *IEEE Trans. Biomed. Eng.*, vol. 68, no. 4, pp. 1123–1130, Apr. 2021, doi: 10.1109/TBME.2020.3011842.

[6] T. Halder, S. Talwar, A. K. Jaiswal, and A. Banerjee, "Performance evaluation of inverse methods for identification and characterization of oscillatory brain sources: Ground truth validation & empirical evidences." bioRxiv, p. 395780, Jan. 17, 2019. doi: 10.1101/395780.

[7] P.-M. RD., "Review of methods for solving the EEG inverse problem," *Int J Bioelectromagn.*, vol. 1, no. 1, pp. 75–86, 1999.

[8] R. D. Pascual-Marqui, C. M. Michel, and D. Lehmann, "Low resolution electromagnetic tomography: a new method for localizing electrical activity in the brain," *Int. J. Psychophysiol.*, vol. 18, no. 1, pp. 49–65, Oct. 1994, doi: 10.1016/0167-8760(84)90014-X.

[9] R. D. Pascual-Marqui, "Standardized low-resolution brain electromagnetic tomography (sLORETA): technical details," *Methods Find. Exp. Clin. Pharmacol.*, vol. 24 Suppl D, pp. 5–12, 2002.

[10] M. A. Jatoi, N. Kamel, A. S. Malik, and I. Faye, "EEG based brain source localization comparison of sLORETA and eLORETA," *Australas. Phys. Eng. Sci. Med.*, vol. 37, no. 4, pp. 713–721, Dec. 2014, doi: 10.1007/s13246-014-0308-3.

[11] S. Asadzadeh, T. Yousefi Rezaii, S. Beheshti, A. Delpak, and S. Meshgini, "A systematic review of EEG source localization techniques

[11] and their applications on diagnosis of brain abnormalities," *J. Neurosci. Methods*, vol. 339, p. 108740, Jun. 2020, doi: 10.1016/j.jneumeth.2020.108740.
[12] J. Vorwerk, Ü. Aydin, C. H. Wolters, and C. R. Butson, "Influence of Head Tissue Conductivity Uncertainties on EEG Dipole Reconstruction," *Front. Neurosci.*, vol. 13, p. 531, 2019, doi: 10.3389/fnins.2019.00531.
[13] O. Steinsträter, S. Sillekens, M. Junghoefer, M. Burger, and C. H. Wolters, "Sensitivity of beamformer source analysis to deficiencies in forward modeling," *Hum. Brain Mapp.*, vol. 31, no. 12, pp. 1907–1927, Dec. 2010, doi: 10.1002/hbm.20986.
[14] H. Becker, J. Fleureau, P. Guillotel, F. Wendling, I. Merlet, and L. Albera, "Emotion Recognition Based on High-Resolution EEG Recordings and Reconstructed Brain Sources," *IEEE Trans. Affect. Comput.*, vol. 11, no. 2, pp. 244–257, Apr. 2020, doi: 10.1109/TAFFC.2017.2768030.
[15] C. J. Holmes, R. Hoge, L. Collins, R. Woods, A. W. Toga, and A. C. Evans, "Enhancement of MR images using registration for signal averaging," *J. Comput. Assist. Tomogr.*, vol. 22, no. 2, pp. 324–333, Apr. 1998, doi: 10.1097/00004728-199803000-00032.
[16] J. C. Mazziotta, A. W. Toga, A. Evans, P. Fox, and J. Lancaster, "A probabilistic atlas of the human brain: theory and rationale for its development. The International Consortium for Brain Mapping (ICBM)," *NeuroImage*, vol. 2, no. 2, pp. 89–101, Jun. 1995, doi: 10.1006/nimg.1995.1012.
[17] Y. Huang, L. C. Parra, and S. Haufe, "The New York Head—A precise standardized volume conductor model for EEG source localization and tES targeting," *NeuroImage*, vol. 140, pp. 150–162, Oct. 2016, doi: 10.1016/j.neuroimage.2015.12.019.
[18] S. D. Fickling *et al.*, "Distant Sensor Prediction of Event-Related Potentials," *IEEE Trans. Biomed. Eng.*, vol. 67, no. 10, pp. 2916–2924, Oct. 2020, doi: 10.1109/TBME.2020.2973617.
[19] L. Zagorchev *et al.*, "Patient-Specific Sensor Registration for Electrical Source Imaging Using a Deformable Head Model," *IEEE Trans. Biomed. Eng.*, vol. 68, no. 1, pp. 267–275, Jan. 2021, doi: 10.1109/TBME.2020.3003112.
[20] S. A. Hossein Hosseini, A. Sohrabpour, M. Akçakaya, and B. He, "Electromagnetic Brain Source Imaging by Means of a Robust Minimum Variance Beamformer," *IEEE Trans. Biomed. Eng.*, vol. 65, no. 10, pp. 2365–2374, Oct. 2018, doi: 10.1109/TBME.2018.2859204.
[21] G. Wang and D. Ren, "Effect of brain-to-skull conductivity ratio on EEG source localization accuracy," *BioMed Res. Int.*, vol. 2013, p. 459346, 2013, doi: 10.1155/2013/459346.
[22] M. Stenroos and O. Hauk, "Minimum-norm cortical source estimation in layered head models is robust against skull conductivity error," *NeuroImage*, vol. 81, pp. 265–272, Nov. 2013, doi: 10.1016/j.neuroimage.2013.04.086.
[23] Z. Akalin Acar and S. Makeig, "Effects of Forward Model Errors on EEG Source Localization," *Brain Topogr.*, vol. 26, no. 3, pp. 378–396, Jul. 2013, doi: 10.1007/s10548-012-0274-6.
[24] Y. Wang and J. Gotman, "The influence of electrode location errors on EEG dipole source localization with a realistic head model," *Clin. Neurophysiol. Off. J. Int. Fed. Clin. Neurophysiol.*, vol. 112, no. 9, pp. 1777–1780, Sep. 2001, doi: 10.1016/s1388-2457(01)00594-6.
[25] J. Gross and A. A. Ioannides, "Linear transformations of data space in MEG," *Phys. Med. Biol.*, vol. 44, no. 8, pp. 2081–2097, Aug. 1999, doi: 10.1088/0031-9155/44/8/317.
[26] K. Sekihara, S. S. Nagarajan, D. Poeppel, A. Marantz, and Y. Miyashita, "Application of an MEG eigenspace beamformer to reconstructing spatio-temporal activities of neural sources," *Hum. Brain Mapp.*, vol. 15, no. 4, pp. 199–215, Apr. 2002, doi: 10.1002/hbm.10019.
[27] S. A. Vorobyov, A. B. Gershman, and Z.-Q. Luo, "Robust adaptive beamforming using worst-case performance optimization: a solution to the signal mismatch problem," *IEEE Trans. Signal Process.*, vol. 51, no. 2, pp. 313–324, Feb. 2003, doi: 10.1109/TSP.2002.806865.
[28] K. Sekihara, S. S. Nagarajan, D. Poeppel, A. Marantz, and Y. Miyashita, "Reconstructing spatio-temporal activities of neural sources using an MEG vector beamformer technique," *IEEE Trans. Biomed. Eng.*, vol. 48, no. 7, pp. 760–771, Jul. 2001, doi: 10.1109/10.930901.
[29] A. Koulouri, V. Rimpiläinen, M. Brookes, and J. P. Kaipio, "Compensation of domain modelling errors in the inverse source problem of the Poisson equation: Application in electroencephalographic imaging," *Appl. Numer. Math.*, vol. 106, pp. 24–36, Aug. 2016, doi: 10.1016/j.apnum.2016.01.005.
[30] V. Rimpiläinen, A. Koulouri, F. Lucka, J. P. Kaipio, and C. H. Wolters, "Improved EEG source localization with Bayesian uncertainty modelling of unknown skull conductivity," *NeuroImage*, vol. 188, pp. 252–260, Mar. 2019, doi: 10.1016/j.neuroimage.2018.11.058.
[31] Z. Akalin Acar, C. E. Acar, and S. Makeig, "Simultaneous head tissue conductivity and EEG source location estimation," *NeuroImage*, vol. 124, pp. 168–180, Jan. 2016, doi: 10.1016/j.neuroimage.2015.08.032.
[32] M.-X. Huang *et al.*, "A novel integrated MEG and EEG analysis method for dipolar sources," *NeuroImage*, vol. 37, no. 3, pp. 731–748, Sep. 2007, doi: 10.1016/j.neuroimage.2007.06.002.
[33] B. D. Van Veen, W. Van Drongelen, M. Yuchtman, and A. Suzuki, "Localization of brain electrical activity via linearly constrained minimum variance spatial filtering," *IEEE Trans. Biomed. Eng.*, vol. 44, no. 9, pp. 867–880, Sep. 1997, doi: 10.1109/10.623056.
[34] R. G. Lorenz and S. P. Boyd, "Robust minimum variance beamforming," *IEEE Trans. Signal Process.*, vol. 53, no. 5, pp. 1684–1696, May 2005, doi: 10.1109/TSP.2005.845436.
[35] P. Chrapka, J. Reilly, and H. de Bruin, "Estimating neural sources using a worst-case robust adaptive beamforming approach," *Biomed. Signal Process. Control*, vol. 52, pp. 330–340, Jul. 2019, doi: 10.1016/j.bspc.2019.04.021.
[36] J. de J. Rubio, "Stability Analysis of the Modified Levenberg–Marquardt Algorithm for the Artificial Neural Network Training," *IEEE Trans. Neural Netw. Learn. Syst.*, vol. 32, no. 8, pp. 3510–3524, Aug. 2021, doi: 10.1109/TNNLS.2020.3015200.
[37] A. Noroozi, R. P. R. Hasanzadeh, and M. Ravan, "A Fuzzy Learning Approach for Identification of Arbitrary Crack Profiles Using ACFM Technique," *IEEE Trans. Magn.*, vol. 49, no. 9, pp. 5016–5027, Sep. 2013, doi: 10.1109/TMAG.2013.2254718.
[38] A. Noroozi, R. P. Hasanzadeh, and M. Ravan, "A fuzzy alignment approach for identification of arbitrary crack shape profiles in metallic structures using ACFM technique," in *20th Iranian Conference on Electrical Engineering (ICEE2012)*, May 2012, pp. 894–899. doi: 10.1109/IranianCEE.2012.6292480.
[39] C. Bishop, "Exact Calculation of the Hessian Matrix for the Multilayer Perceptron," *Neural Comput.*, vol. 4, no. 4, pp. 494–501, Jul. 1992, doi: 10.1162/neco.1992.4.4.494.
[40] R. Oostenveld, P. Fries, E. Maris, and J.-M. Schoffelen, "FieldTrip: Open Source Software for Advanced Analysis of MEG, EEG, and Invasive Electrophysiological Data," *Comput. Intell. Neurosci.*, vol. 2011, p. e156869, Dec. 2010, doi: 10.1155/2011/156869.
[41] J. Vorwerk, R. Oostenveld, M. C. Piastra, L. Magyari, and C. H. Wolters, "The FieldTrip-SimBio pipeline for EEG forward solutions," *Biomed. Eng. Online*, vol. 17, no. 1, p. 37, Mar. 2018, doi: 10.1186/s12938-018-0463-y.
[42] A. Gramfort, T. Papadopoulo, E. Olivi, and M. Clerc, "OpenMEEG: opensource software for quasistatic bioelectromagnetics," *Biomed. Eng. OnLine*, vol. 9, no. 1, p. 45, Sep. 2010, doi: 10.1186/1475-925X-9-45.
[43] A. Sarkar, R. J. Santiago, R. Smith, and A. Kassaee, "Comparison of manual vs. automated multimodality (CT-MRI) image registration for brain tumors," *Med. Dosim. Off. J. Am. Assoc. Med. Dosim.*, vol. 30, no. 1, pp. 20–24, 2005, doi: 10.1016/j.meddos.2004.10.004.
[44] V. Jurcak, D. Tsuzuki, and I. Dan, "10/20, 10/10, and 10/5 systems revisited: their validity as relative head-surface-based positioning systems," *NeuroImage*, vol. 34, no. 4, pp. 1600–1611, Feb. 2007, doi: 10.1016/j.neuroimage.2006.09.024.
[45] L. R. Krol, J. Pawlitzki, F. Lotte, K. Gramann, and T. O. Zander, "SEREEGA: Simulating event-related EEG activity," *J. Neurosci. Methods*, vol. 309, pp. 13–24, Nov. 2018, doi: 10.1016/j.jneumeth.2018.08.001.
[46] X. Mao, G. Marion, and E. Renshaw, "Environmental Brownian noise suppresses explosions in population dynamics," *Stoch. Process. Their Appl.*, vol. 97, no. 1, pp. 95–110, Jan. 2002, doi: 10.1016/S0304-4149(01)00126-0.
[47] H. McCann, G. Pisano, and L. Beltrachini, "Variation in Reported Human Head Tissue Electrical Conductivity Values," *Brain Topogr.*, vol. 32, no. 5, pp. 825–858, Sep. 2019, doi: 10.1007/s10548-019-00710-2.
[48] S. Chabardès *et al.*, "The temporopolar cortex plays a pivotal role in temporal lobe seizures," *Brain J. Neurol.*, vol. 128, no. Pt 8, pp. 1818–1831, Aug. 2005, doi: 10.1093/brain/awh512.
[49] A. Chatzikonstantinou, "Epilepsy and the hippocampus," *Front. Neurol. Neurosci.*, vol. 34, pp. 121–142, 2014, doi: 10.1159/000356435.